\documentclass[12pt]{article}

\usepackage{amssymb}

\usepackage[utf8]{inputenc}
\usepackage[english]{babel}
\usepackage[smaller,nolist]{acronym}
\usepackage{amsmath}
\usepackage{hyperref}

\usepackage{caption}
\usepackage{subcaption}
\usepackage{cleveref}
\usepackage{tabularx}
\usepackage{algorithm2e}
\usepackage{url}
\usepackage{booktabs}
\usepackage{multirow}
\usepackage{float}
\usepackage{graphicx}
\usepackage{makecell}

\DeclareMathOperator*{\argmax}{arg\,max}

\begin{acronym}
    \acro{RS}{Recommender System}
    \acro{CF}{Collaborative Filtering}
    \acro{DL}{Deep Learning}
    \acro{MF}{Matrix Factorization}
    \acro{CF4J}{Collaborative Filtering for Java}
    \acro{ANN}{Artificial Neural Network}
    \acro{KNN}{k-Nearest Neighbors}
    \acro{BeMF}{Bernoulli Matrix Factorization}
    \acro{DirMF}{Dirichlet Matrix Factorization}
    \acro{ResBeMF}{Restricted Bernoulli Matrix Factorization}
    \acro{mAP}{mean Average Precision}
    \acro{PMF}{Probabilistic Matrix Factorization}
    \acro{MLP}{Multi-Layer Perceptron}
    \acro{GCMC}{Graph Convolutional Matrix Completion}
    \acro{MAE}{Mean Absolute Error}
    \acro{MWGP}{Multi-Way Gaussian Processes}
    \acro{BiasedMF}{Biased Matrix Factorization}
    \acro{NGCF}{Neural Graph Collaborative Filtering}
\end{acronym}

\newcommand{\footremember}[2]{%
    \footnote{#2}
    \newcounter{#1}
    \setcounter{#1}{\value{footnote}}%
}
\newcommand{\footrecall}[1]{%
    \footnotemark[\value{#1}]%
} 

\title{Restricted Bernoulli Matrix Factorization: Balancing the trade-off between prediction accuracy and coverage in classification based collaborative filtering}

\author{%
  Ángel González-Prieto\footremember{ucm}{Departamento de Algebra, Geometria y Topologia, Universidad Complutense de Madrid}\footremember{icmat}{Instituto de Ciencias Matematicas (CSIC-UAM-UCM-UC3M)}\footremember{knodis}{KNODIS Research Group, Universidad Politecnica de Madrid}%
  \and Abraham Gutiérrez\footremember{upm}{Departamento de Sistemas Informaticos, Universidad Politecnica de Madrid}\footrecall{knodis}%
  \and Fernando Ortega\footrecall{upm} \footrecall{knodis}%
  \and Raúl Lara-Cabrera\footrecall{upm} \footrecall{knodis}%
  }

\date{}

\begin{document}

\maketitle

\begin{abstract}
Reliability measures associated with the prediction of the machine learning models are critical to strengthening user confidence in artificial intelligence. Therefore, those models that provide not only predictions, but also reliability, enjoy greater popularity. In the field of recommender systems, reliability is crucial, since users tend to prefer those recommendations that are sure to interest them, that is, high predictions with high reliabilities. In this paper, we propose Restricted Bernoulli Matrix Factorization (ResBeMF), a new algorithm aimed at enhancing the performance of classification-based collaborative filtering. This model is based on a collection of restricted matrix factorizations that jointly generate, for each user–item pair, a full probability distribution over the possible rating scores. To prove its effectiveness, the proposed model has been compared to other existing solutions in the literature in terms of prediction quality (Mean Absolute Error and accuracy scores), prediction quantity (coverage score) and recommendation quality (Mean Average Precision score). The experimental results demonstrate that the proposed model provides a good balance in terms of the quality measures used compared to other recommendation models.

\end{abstract}

\section{Introduction} \label{sec:intro}

In the age of the information society, people's daily lives are conditioned by a large number of cloud services. A clear example of this is the way people consume entertainment products. Today, it is difficult to think about watching a movie without Netflix, listening to a song without Spotify, or playing video games without Microsoft Game Pass. However, these types of service have a problem: They offer such a large amount of content to their users that customers often feel lost and disoriented and do not know which item to consume. This is known as the information overload problem~\cite{roetzel2019information}.

\acp{RS}~\cite{resnick1997recommender,park2012literature,bobadilla2013recommender,gorgoglione2019recommendation} are powerful machine learning-based tools that alleviate the problem of information overload. They are also known as filters, since they block information that is irrelevant to users and let pass information that matches the user's preferences. 

The most popular implementation of \acp{RS} is \ac{CF}~\cite{koren2022advances, he2017neural, kluver2018rating, papadakis2022collaborative}. \ac{CF} is a recommendation technique that suggests content to a user by leveraging the preferences or behaviors of others with similar tastes. Its strength lies in needing only the ratings, without requiring any additional information about users or items, allowing it to detect complex patterns and deliver personalized, relevant recommendations. CF algorithms rely on these ratings to make suggestions that align with those of similar users. \Ac{CF} algorithms use these ratings to make recommendations based on the ratings of other users similar to a given one.

Depending on how the recommendations are computed, \ac{CF} is divided into three classes. The first one are \Ac{KNN} based \ac{CF}~\cite{ahn2008new,bobadilla2010new} computes the recommendations by finding the $K$ most similar users of a given one and selecting the preferred items of these users. In a different style, we find \Ac{MF} based \ac{CF}~\cite{mnih2007probabilistic, koren2009matrix}, which issues the recommendations by learning a low-dimensional latent representation of the matrix that contains the ratings. Finally, Neural based \Ac{CF}~\cite{he2017neural,rendle2020neural} creates the recommendations using an \ac{ANN} that extracts the patterns of votes made by users on the items. Graph based models have become very popular in recent days~\cite{wu2022graph,tang2021dynamic,xia2022hypergraph}.

Over the last decade, \ac{CF} has evolved from purely accuracy-driven methods to more expressive probabilistic and uncertainty-aware models. This shift has been motivated by the increasing deployment of recommender systems in real-world settings, where understanding the confidence of predictions is as important as the predictions themselves. As a consequence, recent research has progressively moved beyond point estimates and toward models capable of quantifying uncertainty and reliability in a principled manner. In this context, a growing body of work has highlighted the limitations of evaluating recommender systems solely in terms of predictive accuracy.

For this reason, current trends in \ac{RS} research focus on measuring system performance beyond prediction accuracy~\cite{wu2022survey}, being one of the most popular measures its reliability~\cite{bobadilla2018reliability,katarya2018reliable,ahmadian2019novel}. Reliability indicates the system's certainty in knowing whether a prediction is correct or not. Adding reliability to \acp{RS} helps to improve user confidence in them. On the one hand, if the prediction is correct and the reliability is high, user satisfaction increases. On the other hand, if the prediction fails but the reliability is low, the user can forgive the error. Despite these advances, designing recommendation models that jointly provide accurate predictions, meaningful reliability estimates, and sufficient coverage remains a challenging open problem.

In this paper, we propose a new \ac{CF}-based \ac{MF} model that provides reliable predictions: \ac{ResBeMF}. This model aims to enhance the performance of \ac{CF}-based \ac{RS} that provide reliability linked to their predictions. To achieve this, the \ac{RS} is approached as a multi-objective optimization problem, where the goal is to find a \ac{RS} that provides the best quality of predictions without diminishing the quantity of reliable predictions that the system can offer. \ac{ResBeMF} is motivated by the need for recommender systems that can provide reliable, well-calibrated predictions while maintaining sufficient coverage in practical scenarios. In this regard, \ac{ResBeMF} formulates collaborative filtering as a classification problem and predicts a full probability distribution over rating categories. In this manner, \ac{ResBeMF} enables a principled notion of reliability that is intrinsic to the model rather than imposed a posteriori. The key design choice of introducing structural constraints among rating probabilities allows the model to avoid both overconfident and overly conservative behaviors, such as in prior classification-based models, yielding predictions that are simultaneously interpretable, controllable, and adaptable to different reliability requirements. 

The rest of the document is structured as follows: \Cref{sec:related-work} dives into the related work, \Cref{sec:resbemf} describes the proposed model, \Cref{sec:results} shows the experimental results of the proposed model on gold standard datasets in the field of \ac{CF}, and \cref{sec:conclusions} presents the conclusions and future work of this research.

\section{Related Work} \label{sec:related-work}

Incorporating reliability into \acp{RS} is a hot research topic in recent years. In \cite{katarya2018reliable}, a reliable \ac{RS} is constructed by combining the output of content-based filtering and \ac{CF} using fuzzy cognitive maps. In a different vein, \cite{ahmadian2019novel} presents a novel \ac{RS} which is based on three different points of view on reliability measures: (i) a user-based reliability measure used to evaluate the performance of user rating profiles in predicting unseen items, (ii) an item-based reliability measure used to improve low-quality rating profiles, and (iii) a rating-based reliability measure used to evaluate initial predicted ratings. Furthermore, \cite{zhu2018assigning} proposes a \ac{MF} architecture to provide a reliability value for each prediction in any \ac{CF} based \ac{RS}. These reliability values show improvements in the quality of prediction and recommendation in different \ac{RS}. Recent advancements have further formalised these concepts into the broader framework of Trustworthy \acp{RS}~\cite{li2023trustworthy}. This modern perspective emphasizes that reliability is not just an auxiliary metric, but a core requirement for robustness and user safety in high-stakes decision-making environments.

In addition, the evaluation of reliability measures has also been studied. In \cite{bobadilla2018reliability} two scores are proposed to measure the quality of a reliability measure. Both quality measures are based on the hypothesis that the more suitable a reliability measure is, the better accuracy results it will provide when applied.

The algorithms mentioned above seek to add a quality measure to an existing \ac{CF} model. However, there exists another trend that aims to create \acp{RS} that intrinsically embed reliability into their models. In other words, the output of these models is a tuple $\langle \textrm{prediction}, \textrm{reliability}\rangle$, instead of just a $\textrm{prediction}$ as in regular methods. To this end, these models change the paradigm on which the \ac{RS} algorithms are based. Traditionally, predicting the rating of a user $u$ to an item $i$ has been treated as a regression problem, despite the fact that rating scores are usually not represented by continuous values (e.g., in MovieLens, scores are a discrete set of 1 to 5 stars). These new models that provide both the prediction and its associated reliability treat \ac{RS} as a classification problem.

In this spirit, \ac{BeMF} was proposed in \cite{ortega2021providing}. \Ac{BeMF} is a \ac{CF}-based \ac{RS} that assumes a Bernoulli distribution for the ratings by representing the known ratings using one hot encoding. However, this model assumes that these binary ratings are pairwise independent. More explicitly, suppose that we have a finite set of possible ratings $\mathcal{S}$ (typically, $\mathcal{S}$ is a set of numerical values, such as $\mathcal{S} = \{1, \ldots, 5\}$). Using \ac{MF}, \ac{BeMF} generates, for each user $u$, item $i$, and possible score $s \in \mathcal{S}$, a value $0 \leq \hat{r}_{u,i}^s \leq 1$ to be interpreted as the likelihood that $u$ rates $i$ with a score $s$. However, these values are trained independently for each possible score $s \in \mathcal{S}$ as $|\mathcal{S}|$ different classification problems and, in general, it yields values $\hat{r}_{u,i}^s$ with $\sum_{s} \hat{r}_{u,i}^s \neq 1$, so they do not form a probability distribution on $\mathcal{S}$. In other words, the random variables $\hat{r}_{u,i}^s$ for $s \in \mathcal{S}$ are trained as independent variables. Of course, these values can be normalized to form an a posteriori probability distribution, but this leads to an intrinsic distortion of the results by an abnormal normalization.

To address this issue, \Ac{DirMF}, presented in \cite{lara2022dirichlet}, assumes a Dirichlet distribution for the ratings that avoids the independence of the ratings. However, in this case, for each user $u$ and item $i$, the predicted value $\hat{r}_{u,i}$ is not a probability distribution on the set of ratings $\mathcal{S}$, but rather the hyperparameters of a generator of probability distributions by means of Dirichlet distribution. In other words, $\hat{r}_{u,i}$ is itself a random variable that, when sampled, leads to a probability distribution on $\mathcal{S}$, but only after a second stochastic process.

For these reasons, these classification-based \ac{CF}-methods suffer a dichotomy in their performance that prevents their extension to more general scenarios. In the case of \ac{BeMF}, by design of the model, the rating scores are treated as Bernoulli (binary) independent variables. This is a very unnatural assumption that disregards some subtle patterns and ignores the information provided by the natural ordering in the scores, leading to an output that must be artificially normalized to be interpreted as a probability distribution. This has the effect that the output distributions tend to be very spiky, leading to risky predictions. Hence, if we restrict ourselves to recommend predictions with only high reliability, the system tends to be very aggressive and gets a high coverage but with a low accuracy.

On the other side of the spectrum, \ac{DirMF} does treat the different ratings as dependent variables, and the output of the model can be directly understood as a probability distribution in a very flexible class. But this produces the opposite effect of \ac{BeMF}: the output distributions tend to be very flat. Hence, the system is very conservative and thus, the coverage that we can get with high reliability is very small.

This intrinsic dichotomy reveals a missing middle ground: a model that preserves the expressive power and computational efficiency of matrix factorization while enforcing meaningful dependencies among rating probabilities, thereby achieving a better balance between prediction accuracy and coverage. Addressing this gap is precisely the motivation behind \ac{ResBeMF}, which introduces structural constraints into the Bernoulli framework to couple rating probabilities by design. Concretely, \ac{ResBeMF} is constrained to produce, at each step of the training, a probability distribution $\hat{r}_{u,i}$ on the set of possible ratings $\mathcal{S}$, for each user $u$ and item $i$. In other words, the outcome always satisfies $\sum_{s} \hat{r}_{u,i}^s = 1$ and the random variables $\hat{r}_{u,i}^s$ are treated as dependent for different $s$, and trained simultaneously. This requires applying techniques of constrained optimization during the training phase, and yields to reliable, interpretable predictions without sacrificing practical usability. 

Finally, it is worth mentioning that, beyond these \ac{MF} approaches, the paradigm of modeling probability distributions in recommendations has gained traction through Generative AI, specifically Diffusion Models~\cite{wang2023diffusion}. While these models offer powerful distribution estimation capabilities, they often incur high computational costs during inference. Furthermore, statistical approaches like Conformal Prediction have emerged to provide rigorous guarantees on prediction coverage and reliability~\cite{si2023conformal}. In contrast to these heavy deep learning approaches, our proposed ResBeMF maintains the efficiency of \ac{MF} while intrinsically embedding reliability into the classification process, offering a practical balance between computational scalability and probabilistic confidence.

\section{Restricted Bernoulli Matrix Factorization} \label{sec:resbemf}

The aim of this work is to balance between both \ac{BeMF} and \ac{DirMF} approaches, obtaining a model able to provide accurate predictions with large coverage at the high-reliability regime. For this purpose, we will modify \ac{BeMF} to force the reliability scores to be dependent on design. As we will show, this choice also produces several matrix factorizations, one for each possible rating as in \ac{BeMF}, but the training of each factorization affects the other ones. This leads to a constrained (restricted) optimization problem on the Bernoulli approach, leading to a model that we shall call \ac{ResBeMF}.

To be precise, suppose that we have a set of $U$ users that can rate a collection of $I$ items. Possible ratings that can be assigned form a set $\mathcal{S} = \{s_1, \ldots, s_d\}$ (typically the elements of $\mathcal{S}$ are numerical values such as in the MovieLens dataset $\mathcal{S} = \{1,\ldots, 5\}$). The collection of issued ratings is collected in a $|U| \times |I|$ matrix $R = (r_{u,i})$, where $r_{u,i} \in \mathcal{S}$ is the rating that the user $u$ assigned to the item $i$. As is customary in \acp{RS}, if the user $u$ did not rate the item $i$, we shall denote $r_{u,i} = \bullet$.

The key idea of \ac{ResBeMF} is to create, for each user $u$ and item $i$, a discrete probability distribution with support in $\mathcal{S}$, $p_{u,i}: \mathcal{S} \to [0,1]$, such that $p_{u,i}(s)$ is the probability that $u$ would rate $i$ with the score $s$. To construct this distribution, we shall suppose that, for each $s \in \mathcal{S}$, each user $u$ is associated with a vector $P^s_u \in \mathbb{R}^k$ and each item $i$ is associated with a vector $Q^s_i \in \mathbb{R}^k$, the so-called latent vectors or hidden factors. Here $k$ is a fixed hyper-parameter of the model called the latent dimensionality. We shall gather all these latent vectors into matrices $\mathbf{P} = (P^{s}_{u})_{u,s}$ and $\mathbf{Q} = (Q^{s}_{i})_{i,s}$.

In this way, the probability distribution is given by

\begin{equation}
p_{u,i}(s) = p_{u,i}(s \,|\, \mathbf{P}, \mathbf{Q}) = \sigma_s\left(P_u \cdot Q_i\right).
\end{equation}

Here, $P_u \cdot Q_i = (P_u^{s_1} \cdot Q_i^{s_1}, \ldots, P_u^{s_d} \cdot Q_i^{s_d})$ denotes the list of scalar products of the vectors $P_u^s$ and $Q_i^s$ for $s \in \mathcal{S}$, and $\sigma_s$ is the $s$-th component of the softmax function on $\mathcal{S}$

\begin{equation}
    \sigma_s(x_1, \ldots, x_d)  = \frac{e^{x_s}}{{\sum_{t \in \mathcal{S}} e^{x_t}}}.
\end{equation}

Notice that, since $\sum_{s \in \mathcal{S}} \sigma_s(x) = 1$ for all $x \in \mathbb{R}^d$, the function $p_{u,i}: \mathcal{S} \to [0,1]$ is actually a probability on $\mathcal{S}$. This is in sharp contrast with other \ac{MF}-based models such as \ac{BeMF} \cite{ortega2021providing}.

In order to compute the parameters $\mathbf{P}$ and $\mathbf{Q}$, we can consider the likelihood function

\begin{equation}
    \mathcal{L}(\mathbf{P}, \mathbf{Q}) = \prod_{r_{u,i} \neq \bullet} p_{u,i}(r_{u,i} \,|\, \mathbf{P}, \mathbf{Q}) = \prod_{R_{u,i} \neq \bullet} \sigma_{r_{u,i}}\left(P_u \cdot Q_i\right).
\end{equation}

In this way, the log-likelihood function $\ell(\mathbf{P}, \mathbf{Q}) = \log \mathcal{L}(\mathbf{P}, \mathbf{Q})$ is

\begin{equation}
    \ell(\mathbf{P}, \mathbf{Q}) = \sum_{r_{u,i} \neq \bullet} \log\left(\sigma_{r_{u,i}}\left(P_u \cdot Q_i\right)\right).
\end{equation}

Since we want to maximize the log-likelihood function, we compute the gradient of this function. Recall that the derivative of the softmax function is

\begin{equation}
    \frac{\partial \sigma_s}{\partial x_t} = \sigma_s(\delta_{s,t} - \sigma_t),
\end{equation}

where $\delta_{s,t} = 1$ if $s = t$ and $\delta_{s,t} = 0$ otherwise. Hence, the partial derivative is

\begin{equation}
\begin{aligned}
\frac{\partial}{\partial P_{u_0}^{s_0}} \ell(\mathbf{P}, \mathbf{Q}) & = \sum_{ r_{u_0,i} = s_0} {\left(1-\sigma_{s_0}(P_{u_0} \cdot Q_i)\right)}Q_i^{s_0} \\
    & \;\;-\sum_{r_{u_0,i} \neq s_0, \bullet} {\sigma_{s_0}(P_{u_0} \cdot Q_i)}Q_i^{s_0}.
\end{aligned}
\end{equation}

Analogously for $Q_{i_0}^{s_0}$ we get

\begin{equation}
\begin{aligned}
    \frac{\partial}{\partial Q_{i_0}^{s_0}} \ell(\mathbf{P}, \mathbf{Q}) &= \sum_{ r_{u,i_0} = s_0} {\left(1-\sigma_{s_0}(P_{u} \cdot Q_{i_0})\right)}P_u^{s_0} \\
    & \;\;-\sum_{r_{u,i_0} \neq s_0, \bullet} {\sigma_{s_0}(P_{u} \cdot Q_{i_0})}P_u^{s_0}.
\end{aligned}
\end{equation}

In this manner, we can optimize these parameters through stochastic gradient ascend with learning rate $\eta$ with the update rules for a vote $r_{u,i}=s_0$

\begin{equation}
\begin{aligned}
    & P_{u}^{s_0}  \leftarrow P_{u}^{s_0} + \eta{\left(1-\sigma_{s_0}(P_{u} \cdot Q_{i})\right)}Q_i^{s_0}, \\
    & Q_{i}^{s_0}  \leftarrow Q_{i}^{s_0} + \eta{\left(1-\sigma_{s_0}(P_{u} \cdot Q_{i})\right)}P_u^{s_0},\\
    & P_{u}^{s}  \leftarrow P_{u}^{s} - \eta{\sigma_{s_0}(P_{u} \cdot Q_{i})}Q_i^{s_0}, \textrm{       if } s \neq s_0, \\
    & Q_{i}^{s}  \leftarrow Q_{i}^{s} - \eta{\sigma_{s_0}(P_{u} \cdot Q_{i})}P_u^{s_0}, \textrm{       if } s \neq s_0.
\end{aligned}
\end{equation}

Furthermore, if in addition we consider gaussian priors with zero mean and fixed standard deviation for the parameters $\mathbf{P}$ and $\mathbf{Q}$, we then get that the update rules result (see \cite[Section 2.1]{ortega2021providing})

\begin{equation}
\begin{aligned}
    & P_{u}^{s_0}  \leftarrow P_{u}^{s_0} + \eta\left(\left(1-\sigma_{s_0}(P_{u} \cdot Q_{i})\right)Q_i^{s_0} - \gamma P_{u}^{s_0}\right) , \\
    & Q_{i}^{s_0}  \leftarrow Q_{i}^{s_0} + \eta\left(\left(1-\sigma_{s_0}(P_{u} \cdot Q_{i})\right)P_u^{s_0} - \gamma Q_{i}^{s_0}\right) ,\\
    & P_{u}^{s}  \leftarrow P_{u}^{s} - \eta\left(\sigma_{s_0}(P_{u} \cdot Q_{i})Q_i^{s_0} - \gamma P_{u}^{s}\right) , \textrm{       if } s \neq s_0, \\
    & Q_{i}^{s}  \leftarrow Q_{i}^{s} - \eta\left(\sigma_{s_0}(P_{u} \cdot Q_{i})P_u^{s_0} - \gamma Q_{i}^{s}\right) , \textrm{       if } s \neq s_0.
\end{aligned}
\end{equation}

Here $\gamma$ is another hyper-parameter called the ($L^2$) regularization of the model.

Once the model is trained, the optimal parameters $\mathbf{P}$ and $\mathbf{Q}$ are determined, yielding a probability distribution $p_{u,i}: \mathcal{S} \to [0,1]$ for each user-item pair $<u, i>$. These probability distributions represent the predictions generated by the proposed model. To convert them into predicted ratings $\hat{r}_{u,i}$, we apply the `mode criterion' as follows:

\begin{itemize}
    \item The predicted rating for the user $u$ to the item $i$ is
    \begin{equation}
        \hat{r}_{u,i} = \argmax_{s \in \mathcal{S}} \, p_{u,i}(s) = \argmax_{s \in \mathcal{S}} \, \sigma_s\left(P_u \cdot Q_i\right).    
    \end{equation}
    \item The reliability of the previous prediction is
    \begin{equation}
        \rho_{u,i} =  p_{u,i}\left(\hat{r}_{u,i}\right) = \sigma_{\hat{r}_{u,i}}\left(P_u \cdot Q_i\right).
    \end{equation}
\end{itemize}
Finally, if we fix a threshold $0 \leq \theta \leq 1$ of reliability, we set $\hat{r}_{u,i} = \bullet$ provided that the reliability $\rho_{u,i} < \theta$ (no reliable prediction can be issued).

For example, let us consider a standard 5-star rating scale where $\mathcal{S} = \{1, 2, 3, 4, 5\}$. Suppose the model outputs a probability distribution for a specific user-item pair where the highest probability is associated with the score $4$, such that $p_{u,i} = {0.0, 0.0, 0.07, 0.8, 0.13}$ and $p_{u,i}(4) = 0.8$. According to the mode criterion, the predicted rating is $\hat{r}_{u,i} = 4$ with a reliability of $\rho_{u,i} = 0.8$ (or $80\%$). If the system's threshold is set to $\theta = 0.7$, this prediction is accepted; however, if strictly high reliability is required (e.g., $\theta = 0.9$), the system would discard this prediction ($\hat{r}_{u,i} = \bullet$).

\subsection{Algorithmic implementation of ResBeMF}

A pseudocode implementation of the proposed method is included in Algorithm~\ref{alg:resbemf-code}. The algorithm receives as input the ratings $R$, the scores $\mathcal{S}$ and the model hyper-parameters: the number of latent factors $k$, the number of iterations $m$, the regularization $\gamma$ and the learning rate $\eta$. The algorithm output is made up of the latent factor for users ($\mathbf{P}$) and items ($\mathbf{Q}$). The algorithm contains two main loops: the `for each' loop from lines 3 to 21 is used to update users' factors, and the `for each' loop from lines 22 to 40 allows us to update item factors. These loops can be computed in parallel for each user or item, respectively. Inside these loops, the statements for updating users and items factors are equivalent: first, the gradient of all known ratings of the user or item is accumulated in $\Delta$ (users) or $\Phi$ (items), and then the factors are updated proportionally to these gradients and the learning rate $\eta$.

\small
\begin{algorithm}[H]
	\DontPrintSemicolon
	\LinesNumbered
	
    \SetKwInOut{Input}{input}
    \SetKwInOut{Output}{output}

    \Input{$R, k, \gamma, \eta, m, \mathcal{S}$}
	\Output{$\mathbf{P}, \mathbf{Q}$}

	Initialize $\mathbf{P}\leftarrow U(0,1), \mathbf{Q}\leftarrow U(0,1)$\;

	\Repeat{$m$ iterations} {
	    \For{each user $u$}{
	        Initialize $\Delta$ to 0\;
	        
	        \For{each item $i$ rated by user $u$: $R_{u,i}$}{
	            \For{each score $s \in \mathcal{S} = \left \{ s_1, \ldots, s_D \right \}$} {
	                \For{each $f \in \{1, \ldots, k\}$}{
	                    \uIf{$R_{u,i}=s$}{
                            $\Delta_f^s \leftarrow \Delta_f^s + (1-\sigma_{R_{u,i}}(P_u \cdot Q_i)) \cdot Q_{i,f}^s - \gamma \cdot P_{u,f}^s$ \;
                        }
                        \Else{
                            $\Delta_f^s \leftarrow \Delta_f^s - \sigma_{R_{u,i}}(P_u \cdot Q_i) \cdot Q_{i,f}^s - \gamma \cdot P_{u,f}^s$ \;
                        }
	                }
	            }
	        }
	        
	        \For{each score $s \in \mathcal{S} = \left \{ s_1, \ldots, s_D \right \}$} {
	           \For{each $f \in \{1, \ldots, k\}$}{
	               $P_{u,f}^s \leftarrow P_{u,f}^s + \eta \cdot \Delta_f^s$ \;
	            }
	       }
        }
        
        \For{each item $i$}{
            Initialize $\Phi$ to 0\;

	        \For{each user $u$ that rated item $i$: $R_{u,i}$}{
	            \For{each score $s \in \mathcal{S} = \left \{ s_1, \ldots, s_D \right \}$} {
	                \For{each $f \in \{1, \ldots, k\}$}{
	                    \uIf{$R_{u,i}=s$}{
                            $\Phi_f^s \leftarrow \Phi_f^s + (1-\sigma_{R_{u,i}}(P_u \cdot Q_i)) \cdot P_{u,f}^s - \gamma \cdot Q_{i,f}^s$ \;
                        }
                        \Else{
                            $\Phi_f^s \leftarrow \Phi_f^s - \sigma_{R_{u,i}}(P_u \cdot Q_i) \cdot P_{u,f}^s - \gamma \cdot Q_{i,f}^s$ \;
                        }
	                }
	            }
	        }
	        
	        \For{each score $s \in \mathcal{S} = \left \{ s_1, \ldots, s_D \right \}$} {
	           \For{each $f \in \{1, \ldots, k\}$}{
	               $Q_{i,f}^s \leftarrow Q_{i,f}^s + \eta \cdot \Phi_f^s$  \;
	            }
	       }
        }
    }
 
	\caption{ResBeMF model-fitting algorithm}
    \label{alg:resbemf-code}
\end{algorithm}
\normalsize

\section{Experimental results} \label{sec:results}

This section contains the experimental results carried out to evaluate the proposed \ac{ResBeMF} model. All experiments have been performed using the \ac{CF4J}~\cite{ortega2018cf4j, ortega2021cf4j} framework and its source code can be found on GitHub\footnote{\url{https://github.com/KNODIS-Research-Group/resbemf}}.

Experimental evaluation has been conducted using the MovieLens~\cite{harper2015movielens}, FilmTrust~\cite{guo2013novel} and MyAnimeList datasets. The main parameters of these datasets are shown in \cref{tab:datasets}. To facilitate the reproducibility of these experimental results, the default train/test splits of these datasets included in \ac{CF4J} were used.

\begin{table}[H]
\centering
\small
\begin{tabularx}{\textwidth}{|l|X|X|X|X|X|}
\hline
\textbf{Dataset}        & \textbf{Number of users} & \textbf{Number of items} & \textbf{Number of ratings} & \textbf{Number of test ratings} & \textbf{Scores}     \\ 
\hline
FilmTrust      & 1,508           & 2,071           & 32,675            & 2,819                  & 0.5 to 4.0 \\ 
\hline
MovieLens 100K & 943             & 1,682           & 92,026            & 7,974                  & 1 to 5     \\ 
\hline
MovieLens 1M   & 6,040           & 3,706           & 911,031           & 89,178                 & 1 to 5     \\ 
\hline
MovieLens 10M  & 69,878          & 10,677          & 9,104,681	       & 895,373                & 0.5 to 5.0 \\ 
\hline
MyAnimeList    & 69,600          & 9,927           & 5,788,207         & 549,027                & 1 to 10    \\ 
\hline
\end{tabularx}
\caption{Main parameters of the datasets used in the experiments.}
\label{tab:datasets}
\end{table}

During the experiments performed in this section, several baselines have been included to compare the proposed \ac{ResBeMF} method with other approaches. The selection of these baselines has been made trying to ensure that the two current trends in \ac{CF}, \ac{MF}, and \ac{ANN} are represented. Regarding \ac{MF}-based models, we have selected our previous works \ac{BeMF}~\cite{ortega2021providing} and \ac{DirMF}~\cite{lara2022dirichlet}, as, like the proposed method, they are \ac{CF} methods based on classification. Furthermore, we have included \ac{PMF}~\cite{mnih2007probabilistic} and \ac{BiasedMF}~\cite{koren2009matrix} as they are the most popular implementations of \ac{MF}-based \ac{CF}. For the models based on \ac{ANN}, we have chosen \ac{MLP}~\cite{he2017neural}, \ac{GCMC}~\cite{berg2017graph}, \ac{MWGP}~\cite{yang2021multi} and \ac{NGCF}~\cite{wang2019neural}: \ac{MLP} is a simple model that aims to mimic the functioning of \ac{MF} using \ac{ANN}; \ac{GCMC} is a much more complex model that addresses the \ac{CF} problem from the perspective of link prediction on graphs; \ac{MWGP} is an innovative approach that merges the representation learning paradigm of \ac{CF} with multi-output Gaussian processes within a unified framework to produce recommendations that are aware of uncertainty; and \ac{NGCF} explicitly injects collaborative signals into the recommendation process by propagating embeddings across the user-item graph to model high-order connectivity.

\subsection{Quality measures}\label{sec:quality-measures}

As we stated previously, the output of the \ac{ResBeMF} model consists of a discrete probability distribution that indicates the probability that a user $u$ will rate an item $i$ with any of the scores $\mathcal{S}$. This output allows for tuning the degree of reliability of the model by filtering out those predictions that are unreliable. If a high-reliability threshold is set, the model will be able to compute only a few predictions, but these will be very accurate. If a low-reliability threshold is set, the model will be able to compute many predictions, but these will be less accurate. Therefore, the quality of the model should be measured in terms of two quality measures: the error of the predictions and the ability to compute predictions. Note that the reliability threshold is not a static hyperparameter, but a modulation tool that allows system designers to shift the operating point towards either reliability or coverage, depending on the specific application requirements (e.g., risk-averse vs. discovery-oriented systems).

To measure the error of predictions, we can follow two approaches: Consider the problem of predicting the vote as a regression problem, in which case we should use a regression score such as \ac{MAE}, or consider the problem of predicting the vote as a classification problem, in which case we should use a classification score such as accuracy. Both quality measures offer different perspectives on the performance of the models, and therefore, we will use both. For the sake of completeness, we will briefly review these quality measures, as well as their particular incarnations for our model. For a thorough revision of these metrics, please refer to \cite{zhu2004recall}.

We define the \ac{MAE} of the predictions with reliability greater than or equal to a parameter $0 \leq \theta \leq 1$ as

\begin{equation}
    \textrm{MAE}^\theta = \frac{1}{|U|} \sum_{u \in U} \frac{1}{| \hat{R}_u^\theta |} \sum_{i \in \hat{R}_u^\theta} \frac{\mid  r_{u,i} - \hat{r}_{u,i} \mid}{\max (\mathcal{S}) - \min (\mathcal{S})},
\end{equation}

where $U$ is the set of users and $\hat{R}_u^\theta$ is the set of items rated by the user $u$ in the test split with reliability greater than  or equal to $\theta$.

Analogously, we define the accuracy with reliability greater than or equal to $\theta$ as

\begin{equation}
    \textrm{accuracy}^\theta = \frac{1}{|U|} \sum_{u \in U} \frac{1}{| \hat{R}_u^\theta |} \sum_{i \in \hat{R}_u^\theta} \delta \left(r_{u,i},\hat{r}_{u,i}\right),
\end{equation}

where $\delta \left(r_{u,i},\hat{r}_{u,i}\right)$ is an indicator function equal to $1$ if the test rating $r_{u,i}$ is equal to the prediction $\hat{r}_{u,i}$ and $0$ otherwise.

To measure the capability to compute predictions at threshold $\theta$, we use

\begin{equation}
    \textrm{coverage}^\theta = \frac{1}{|U|} \sum_{u \in U} \frac{| \hat{R}_u^\theta |}{| \hat{R}_u |},
\end{equation}

where $\hat{R}_u$ is the whole set of items rated by the user $u$ in the test split.

In any case, the main purpose of a \ac{RS} is to suggest a list of items that may be of interest to a user. Knowing the quality of these lists is crucial to assess the real performance of a \ac{RS}. To measure the quality of the recommendation lists, we use \ac{mAP}. Before defining \ac{mAP}, we define the precision for user $u$ with threshold $\theta$ at $N$ best items to be

\begin{equation}
    \textrm{precision}_u^\theta@N = \frac{\#\{i \in T(N)_u \,|\, r_{u,i} \geq \lambda \}}{N}
\end{equation}

where $T(N)_u^\theta$ is the list of top $N$ items recommended to the user $u$ based on the predictions and $\lambda$ s a threshold that discerns whether an item $i$ is of interest to the user $u$ or not, based on their test rating$r_{u,i}$. Generally, the list of recommendations $T(N)_u^\theta$ is built by selecting the recommended items with the highest predictions. However, when using classification-based \ac{CF}, such as \ac{ResBeMF}, there are many identical predictions, leading to many ties that make it difficult to determine which predictions are the highest. Therefore, in these cases, the sorting is done first based on the mode of the probability distribution generated by the classification based \ac{CF}, and in case of a tie, based on the mean of this probability distribution.

We also define the average precision with threshold $\theta$ at $N$ items to be

\footnotesize
\begin{equation}
    \textrm{AP}_u^\theta@N = \frac{1}{\#\{i \in T(N)_u^\theta \,|\, r_{u,i} \geq \lambda \}} \sum_{k=1}^N \textrm{precision}_u^\theta@N \cdot \textrm{rel}_\lambda(k),
\end{equation}
\normalsize
where $\textrm{rel}_\lambda(k)$ is an indicator function equal to $1$ if the recommended item at rank $k$ is relevant (i.e.\ if its test rating is greater than or equal to $\lambda$), and $0$ otherwise.

And finally we define $\textrm{mAP}$ to be the mean of $\textrm{AP}_u^\theta@N$ among all the users, that is

\begin{equation}
    \textrm{mAP}^\theta@N = \frac{1}{|U|} \sum_{u \in U} \textrm{AP}^\theta_{u}@N.
\end{equation}

\subsection{Multi-objective optimization through hyper-parameter tuning}\label{sec:multiobjective}

At this point, observe that the prediction task with reliability is a multi-objective problem: increasing prediction quality reduces prediction capability and vice versa. Therefore, hyper-parameter tuning should be approached from a multi-objective perspective. In other words, there is no single combination of hyper-parameters that yields the best performance on a given dataset. Instead, multiple solutions, known as the Pareto front, provide models optimized in terms of the balance between prediction quality and quantity.

For this purpose, it is necessary to extend the quality measures defined in \cref{sec:quality-measures}, as these measures report the quality of the model for a fixed reliability threshold $\theta$. To evaluate the real quality of the model, we must average these measures for different values of $\theta$. As $\theta \in [0,1]$, we sample $\theta$ in an equidistant partition of the unit interval with $N$ points

\begin{equation}
    \theta_k = \frac{k}{N-1}
\end{equation}

for $k = 0, \ldots, N-1$. For example, if $N=5$ we have $\theta_0 = 0, \theta_1 = 0.25, \theta_2 = 0.50, \theta_3 = 0.75$ and $\theta_4 = 1.00$. In our experiments, we have fixed $N=20$. To average the results, we define

\begin{equation}
    1-\textrm{MAE} = \frac{2}{(N+1)N} \sum_{k=0}^{N-1} \left(1-\textrm{MAE}^{\theta_k}\right)
\end{equation}

and

\begin{equation}
    \textrm{coverage} = \frac{2}{(N+1) N} \sum_{k=0}^{N-1} \textrm{coverage}^{\theta_{k}}
\end{equation}

The tuning of hyper-parameters was carried out by a random search using 5-fold cross-validation. The hyper-parameters tested are listed in \cref{tab:random-search-hyperparameters}. This random search has been carried out for the three algorithms of identical nature evaluated in this study: \ac{ResBeMF}, \ac{BeMF}~\cite{ortega2021providing}, and \ac{DirMF}~\cite{lara2022dirichlet}. It should be noted that these three methods are \ac{MF} based \ac{CF}, they output a discrete probability distribution, and they are based on gradient descent. For this reason, the hyper-parameters of these models are the same: number of factors, regularization, learning rate, and number of iterations. The main difference between these three algorithms lies in the probability distribution underlying the \ac{RS} ratings.

\begin{table}[H]
\centering
\footnotesize
\setlength{\tabcolsep}{1.0pt}
\begin{tabularx}{\textwidth}{|l|l|*{5}{>{\centering\arraybackslash}X|}}
\hline
\textbf{Model} & \makecell{\textbf{Hyper}\\\textbf{parameter}} & \makecell{\textbf{MovieLens}\\\textbf{100K}} & \makecell{\textbf{MovieLens}\\\textbf{1M}} & \textbf{FilmTrust} & \makecell{\textbf{MovieLens}\\\textbf{10M}} & \textbf{MyAnimeList} \\ \hline
\multirow{4}{*}{ResBeMF} 
 & Number of factors & \multicolumn{3}{c|}{2, 4, 6, 8, 10} & 10, 20, 30, 40 & 2, 4, 6, 8, 10 \\ \cline{2-7} 
 & Regularization    & \multicolumn{4}{c|}{0.01, 0.05, 0.10, 0.15, 0.20} & 0.0001, 0.001, 0.01, 0.1 \\ \cline{2-7} 
 & Learning rate     & \multicolumn{3}{c|}{0.001, 0.002, 0.003, 0.004, 0.005} & \multicolumn{2}{c|}{\makecell{0.0001, 0.0002, 0.0003, \\ 0.0004, 0.0005}} \\ \cline{2-7} 
 & Number of iters   & \multicolumn{5}{c|}{25, 50, 75, 100} \\ \hline
\multirow{4}{*}{BeMF}    
 & Number of factors & \multicolumn{5}{c|}{2, 4, 6, 8, 10} \\ \cline{2-7} 
 & Regularization    & \multicolumn{5}{c|}{0.0001, 0.001, 0.01, 0.1, 1.0} \\ \cline{2-7} 
 & Learning rate     & \multicolumn{3}{c|}{0.01, 0.02, 0.03, 0.04, 0.05} & \multicolumn{2}{c|}{\makecell{0.001, 0.002, 0.003, \\ 0.004, 0.005}} \\ \cline{2-7} 
 & Number of iters   & \multicolumn{5}{c|}{25, 50, 75, 100} \\ \hline
\multirow{4}{*}{DirMF}   
 & Number of factors & \multicolumn{5}{c|}{2, 4, 6, 8, 10} \\ \cline{2-7} 
 & Regularization    & \multicolumn{5}{c|}{0.0001, 0.00025, 0.0005, 0.00075, 0.001} \\ \cline{2-7} 
 & Learning rate     & \multicolumn{5}{c|}{0.1, 0.2, 0.3, 0.4, 0.5} \\ \cline{2-7} 
 & Number of iters   & \multicolumn{5}{c|}{25, 50, 75, 100} \\ \hline
\multicolumn{2}{|l|}{Random Search coverage} & \multicolumn{3}{c|}{1.0} & 0.50 & 0.75 \\ \hline
\end{tabularx}
\caption{Hyper-parameters evaluated during the random search.}
\label{tab:random-search-hyperparameters}
\end{table}

\begin{figure}[H]
    \centering
    \includegraphics[width=0.95\textwidth]{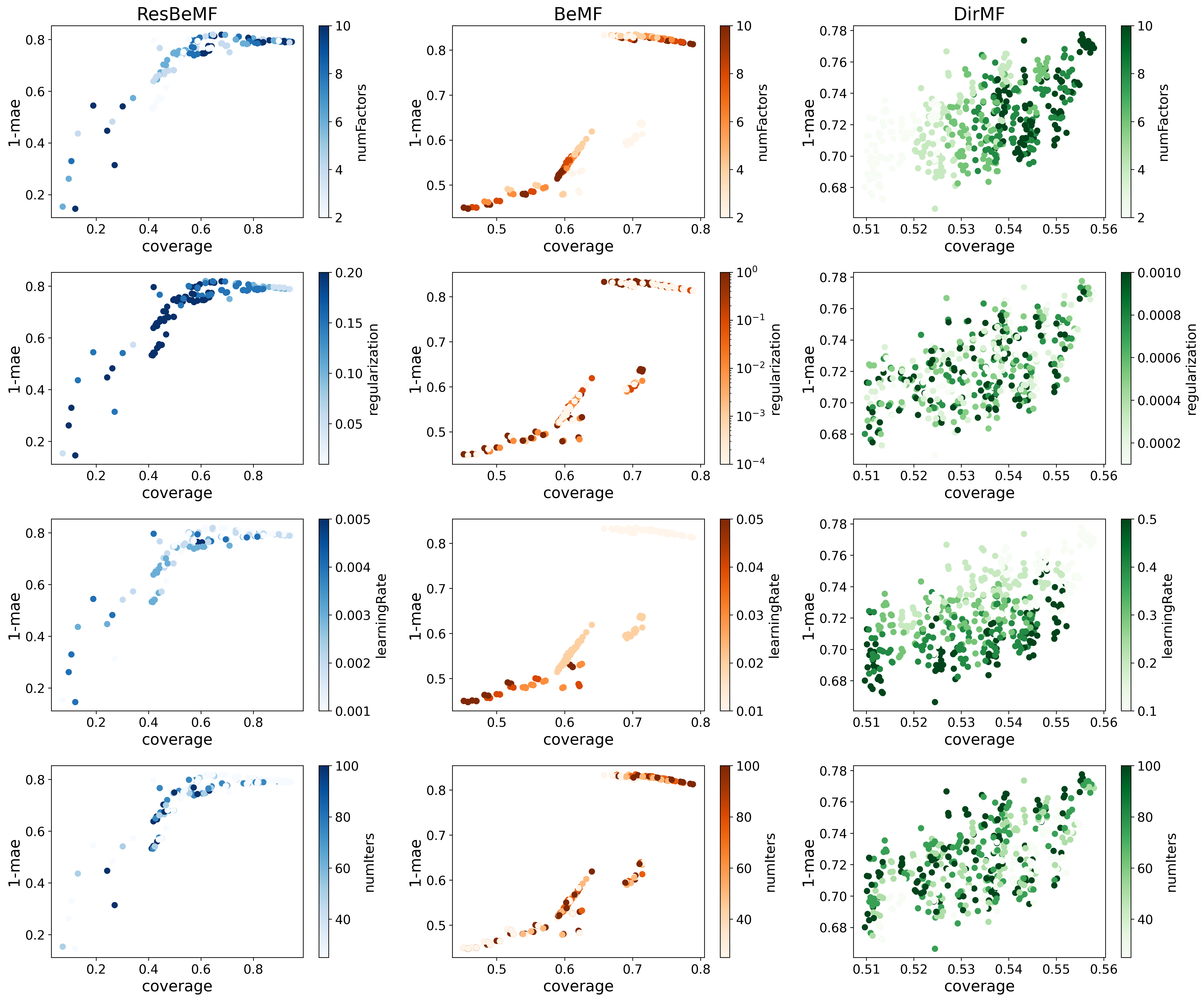}
    \caption{Random search results for parameter tuning in MovieLens 1M dataset (larger is better in both axes).}
    \label{fig:ml1m-hyperparameters}
\end{figure}

\Cref{fig:ml1m-hyperparameters} contains the results of this random search on the MovieLens 1M dataset. The figure shows a total of 12 scatter plots: one plot for each of the 4 hyper-parameters of the 3 models evaluated. This figure shows both the Pareto fronts of each model and the influence of each hyper-parameter on these Pareto fronts. We can observe that: 
\begin{itemize}
    \item The regularization is crucial for \ac{ResBeMF}, since higher values result in accurate predictions but low prediction capability.
    \item The learning rate has a great influence for \ac{BeMF}: Better models are obtained when the learning rate is low.
    \item The number of latent factors modulates the recommendation capability of \ac{DirMF}.
\end{itemize}

The results of the same experiment for the MovieLens 100K, FilmTrust, MyAnimeList and MovieLens 10M datasets can be found in \cref{sec:appendix}.

\begin{figure}[H]
    \centering
    \includegraphics[width=\textwidth]{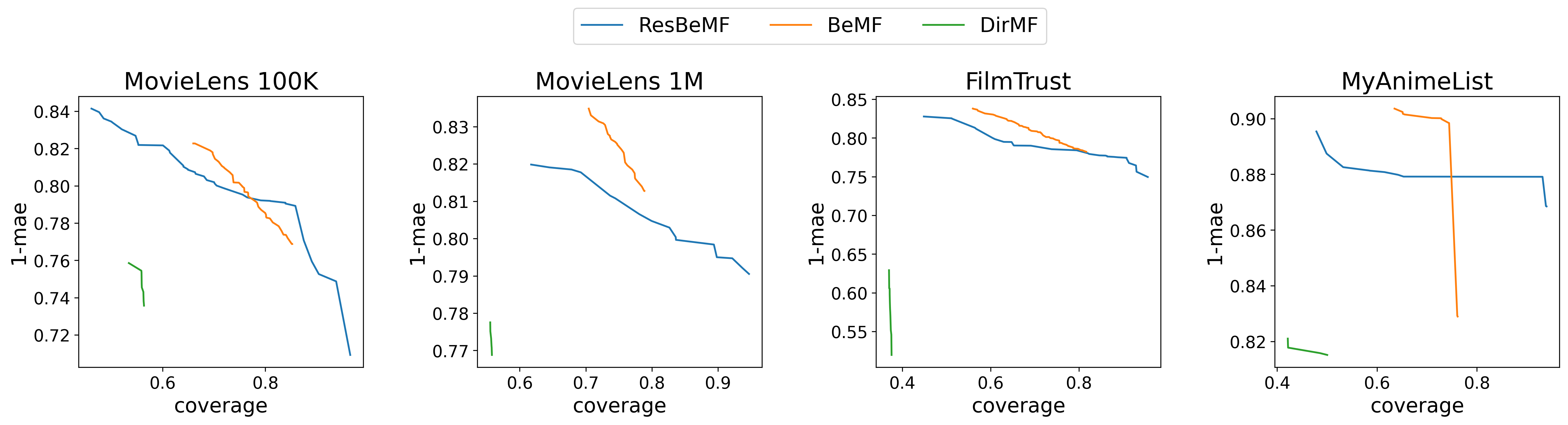}
    \caption{Pareto front comparison for the evaluated datasets (larger is better).}
    \label{fig:pareto-front}
\end{figure}

Using the result of this search, the Pareto front of the validation error of each model can be obtained. \Cref{fig:pareto-front} compares the Pareto fronts in the five datasets evaluated.

We can observe that the proposed \ac{ResBeMF} model provides the best balance between prediction quality and prediction capability across all evaluated datasets. This model is the only one that can cover a wide range of coverage values (from 0.5 to 1.0) without reducing the quality of the predictions. On the contrary, \ac{BeMF} usually provides the highest prediction quality, but its prediction capacity is lower than \ac{ResBeMF} since its range of coverage values is much narrower (from 0.60 to 0.85 in the best case). Finally, \ac{DirMF} usually gives poor results in terms of both prediction quality and prediction capability.

\subsection{Test performance}

The previous experiments show the good performance of the proposed model, \ac{ResBeMF}, using validation data. However, these results are insufficient to conclude that the proposed model is better than the models it has been compared with. First, it is necessary to ensure the absence of model overfitting. To do this, the model should be evaluated on the test partition. This evaluation will be carried out using the Pareto front solutions obtained from the 5 cross-validation splits (i.e., the combination of hyper-parameters that forms the Pareto fronts shown in \cref{fig:pareto-front}). Additionally, it is necessary to compare the model with models of a different nature as well as with other quality measures.

\begin{figure}[H]
    \centering
    \includegraphics[width=0.95\textwidth]{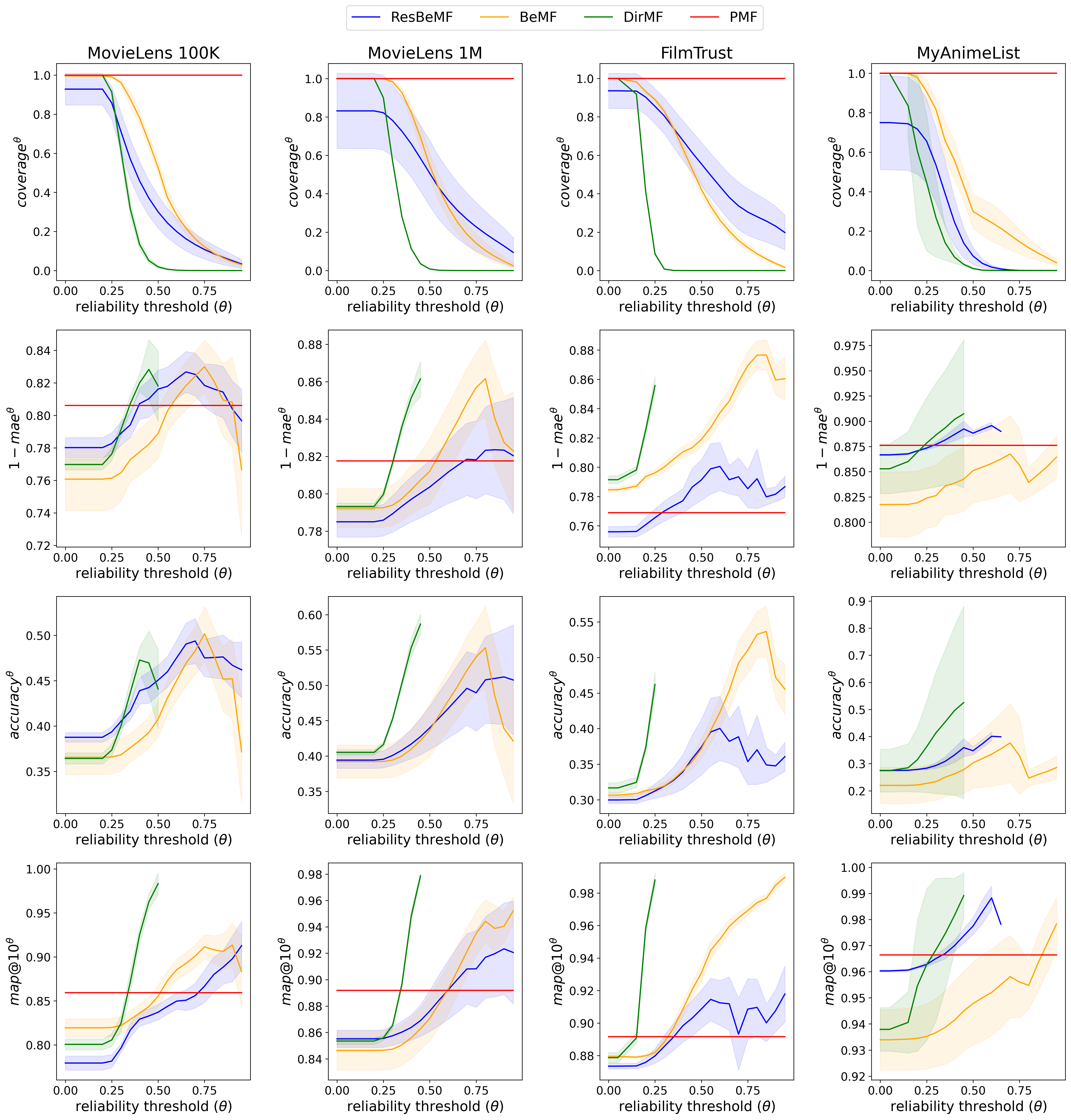}
    \caption{Test error of \ac{MF} based \ac{CF} models.}
    \label{fig:mf-test-error}
\end{figure}

\Cref{fig:mf-test-error} compares the test scores of the proposed method with respect to other \ac{MF} based \ac{CF} methods: \ac{BeMF}, \ac{DirMF}, \ac{PMF} and \ac{BiasedMF} in the selected datasets (see \cref{tab:datasets}). In this experiment, two new quality measures have been added: $\textrm{accuracy}$, available only for classification-based models, and $\textrm{mAP}$. The \ac{ResBeMF}, \ac{BeMF}, and \ac{DirMF} models are represented with a solid line and a shaded area. The solid line shows the average test error across all solutions on the Pareto front, while the shaded area represents the 95\% confidence interval of these errors. In contrast, \ac{PMF} and \ac{BiasedMF} are represented using only a solid line, as these models guarantee a coverage of 1.0 but cannot provide reliability for their predictions. Therefore, hyper-parameter optimization for these models was conducted using a random search aimed at minimizing the \ac{MAE} of the predictions. \Cref{tab:pmf-hyperparameters} and \cref{tab:biasedmf-hyperparameters} contain these hyper-parameters for \ac{PMF} and \ac{BiasedMF} models respectively.

On the one hand we can observe that the general trend of all models is as expected: as the reliability threshold increases, the quantity of predictions decreases while their quality improves. However, it should be noted that, with very high reliability thresholds, the coverage is so low that some anomalies are observed, such as an increase in prediction error. On the other hand we can observe that \ac{DirMF} is the model that provides the best prediction ($\textrm{MAE}$ and $\textrm{accuracy}$) and recommendation ($\textrm{mAP}$) quality in absolute terms, but its prediction capacity is very low; \ac{ResBeMF} has the best prediction and recommendation capacity without degrading the quality of the predictions; \ac{BeMF} is the most balanced model, offering good predictions and recommendation with acceptable coverage in 4 of the 5 evaluated datasets; and \ac{PMF} and \ac{BiasedMF} have neither the best prediction quality nor the best recommendation quality, although their prediction capacity is perfect (recall that no filtering is applied). This reinforces the observation that \ac{ResBeMF} is the \ac{MF} based model that is the most flexible and customizable model, capable of compelling operations in scenarios of high reliability and high accuracy (and thus low coverage), and in scenarios of low reliability and low accuracy (but in contrast, high coverage).

\begin{figure}[H]
    \centering
    \includegraphics[width=0.95\textwidth]{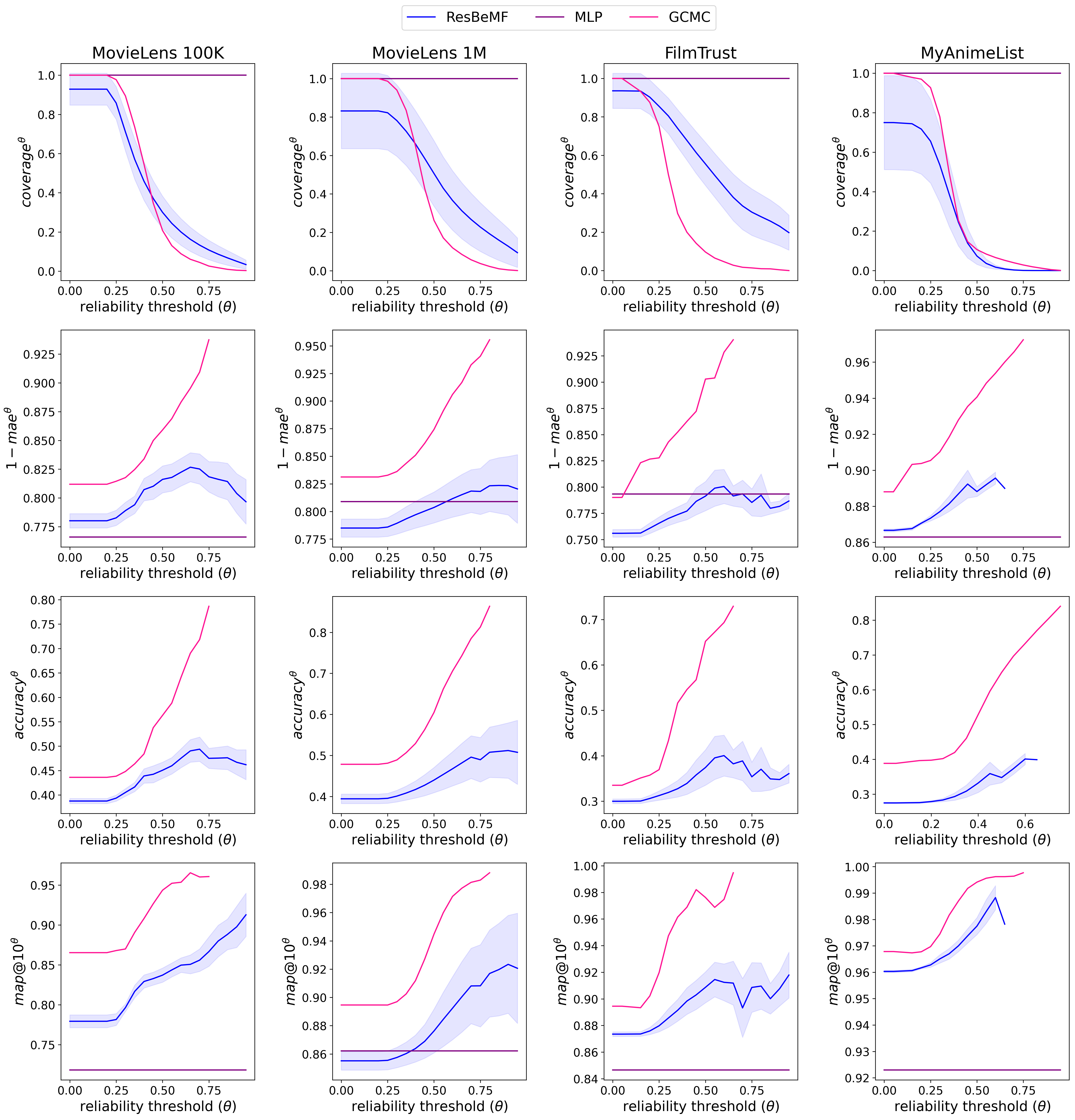}
    \caption{Test error of \ac{ANN} based \ac{CF} models.}
    \label{fig:nn-test-error}
\end{figure}

\Cref{fig:nn-test-error} compares the test scores of the proposed method against other \ac{ANN}-based \ac{CF} methods: \ac{MLP}, \ac{GCMC}, \ac{MWGP} and \ac{NGCF}. In this comparison, all \ac{ANN}-based methods are represented as a solid line without a shaded area. On the one hand, \ac{MLP} does not show a shaded area because, like \ac{PMF} and \ac{BiasedMF}, it is a model that does not provide reliabilities for the predictions. Therefore, hyper-parameter optimization for this model was performed using a random search to minimize the \ac{MAE} of the predictions. The optimal hyper-parameters for each dataset are detailed in \cref{tab:mlp-hyperparameters}. On the other hand, \ac{GCMC} and \ac{NGCF} do provide reliabilities; however, the enormous computational cost of its training makes it impossible to search for different combinations of hyper-parameters that offer a wide Pareto Front, and therefore, it was decided to select the hyper-parameters proposed in their public implementations\footnote{\ac{GCMC}: \url{https://github.com/riannevdberg/gc-mc} \\ \ac{NGCF}: \url{https://github.com/xiangwang1223/neural_graph_collaborative_filtering}}. Finally, \ac{MWGP}, like \ac{MLP}, does not provide reliabilities, so only one set of hyper-parameters was fixed for its execution. Due to the computational cost of its execution, it was decided to choose the hyper-parameters proposed in its public implementation\footnote{\url{https://github.com/Tuyki/mogp-decomposition}}.

\begin{table}[H]
    \centering
    \small
    \begin{subtable}[h]{\textwidth}
        \centering
        \begin{tabular}{|c|c|c|c|c|c|}
        \hline
        \textbf{Hyper parameter} & \makecell{\textbf{MovieLens}\\\textbf{100K}} & \makecell{\textbf{MovieLens}\\\textbf{1M}} & \textbf{FilmTrust} & \makecell{\textbf{MovieLens}\\\textbf{10M}} & \textbf{MyAnimeList} \\ 
        \hline
        Number of factors & 6              & 10           & 2         & 10            & 6           \\ 
        \hline
        $\gamma$          & 0.01           & 0.01         & 0.01      & 0.01          & 0.01        \\ 
        \hline
        $\lambda$         & 0.1            & 0.1          & 0.1       & 0.1           & 0.1         \\ 
        \hline
        Number of iters   & 100            & 100          & 100       & 100           & 100         \\ 
        \hline
        \end{tabular}
       \caption{PMF}
       \label{tab:pmf-hyperparameters}
    \end{subtable}
    \par\bigskip
    \begin{subtable}[h]{\textwidth}
        \centering
        \begin{tabular}{|c|c|c|c|c|c|}
        \hline
        \textbf{Hyper parameter} & \makecell{\textbf{MovieLens}\\\textbf{100K}} & \makecell{\textbf{MovieLens}\\\textbf{1M}} & \textbf{FilmTrust} & \makecell{\textbf{MovieLens}\\\textbf{10M}} & \textbf{MyAnimeList} \\ 
        \hline
        Number of factors & 2              & 10           & 2         & 10            & 8           \\ 
        \hline
        $\gamma$          & 0.01           & 0.01         & 0.01      & 0.01          & 0.01        \\ 
        \hline
        $\lambda$         & 0.1            & 0.1          & 0.1       & 0.1           & 0.1         \\ 
        \hline
        Number of iters   & 100            & 100          & 100       & 100           & 75          \\ 
        \hline
        \end{tabular}
       \caption{BiasedMF}
       \label{tab:biasedmf-hyperparameters}
    \end{subtable}
    \par\bigskip
    \begin{subtable}[h]{\textwidth}
        \centering
        \begin{tabular}{|c|c|c|c|c|c|}
        \hline
        \textbf{Hyper parameter} & \makecell{\textbf{MovieLens}\\\textbf{100K}} & \makecell{\textbf{MovieLens}\\\textbf{1M}} & \textbf{FilmTrust} & \makecell{\textbf{MovieLens}\\\textbf{10M}} & \textbf{MyAnimeList} \\          
        \hline
        Number of factors & 2              & 2            & 8         & 6             & 8           \\ 
        \hline
        Epochs            & 100            & 100          & 100       & 75            & 75          \\ 
        \hline
        Learning rate     & 0.10           & 0.10         & 0.01      & 0.10          & 0.10        \\ 
        \hline
        \end{tabular}
        \caption{MLP}
        \label{tab:mlp-hyperparameters}
     \end{subtable}
     \caption{Best hyper-parameters found using 5-fold cross-validation to minimize the MAE of the predictions.}
     \label{tab:baselines-hyperparameters}
\end{table}

In the results, we can observe a trend similar to that found in the comparison of \ac{ResBeMF} with other \ac{MF} models. On the one hand, the proposed model has a lower coverage than \ac{MLP} and \ac{MWGP}, but it improves the quality of predictions and recommendations. On the other hand, \ac{ResBeMF} has better coverage than \ac{GCMC} and \ac{NGCF}, but the quality of its predictions and recommendations is lower.

\subsection{Scalability}

The proposed model adds new constraints compared to other \ac{MF}-based \ac{CF} models. Although these constraints have been shown to improve prediction and recommendation quality, they may not be useful if they significantly reduce model scalability. \Cref{tab:fitting-times} shows the fitting times in seconds for the \ac{MF}-based models. For models that provide an estimate of the reliability of their predictions, the average time (and standard deviation) required to tune the combinations of hyper-parameters from the Pareto fronts obtained in validation is presented. For models that do not include reliability estimates, the tuning time for the best hyper-parameter combination obtained in validation is shown (see \cref{tab:baselines-hyperparameters}). \ac{ANN}-based models have been excluded from this comparison because they were not run under the same conditions as the MF-based models: \ac{MLP} was implemented using ND4J instead of CF4J; and \ac{GCMC}, \ac{MWGP} and \ac{NGCF} were implemented with PyTorch (Python) instead of \ac{CF4J} (Java) and ran on GPU instead of CPU.

\begin{table}[H]
\centering
\small
\begin{tabular}{|l|c|c|c|c|c|}
\hline
\textbf{Dataset}        & \textbf{ResBeMF}         & \textbf{BeMF}          & \textbf{DirMF}          & \textbf{PMF}  & \textbf{BiasedMF} \\ \hline
MovieLens 100K & 1.54 (0.78)     & 14.07 (2.19)  & 4.52 (0.61)    & 3.24 & 3.25     \\ \hline
MovieLens 1M   & 2.12 (0.58)     & 11.76 (1.78)  & 17.57 (5.24)   & 3.68 & 3.90     \\ \hline
FilmTrust      & 1.43 (0.92)     & 18.43 (6.31)  & 3.07 (0.32)    & 3.10 & 3.14     \\ \hline
MyAnimeList    & 58.71 (25.12)   & 69.91 (14.86) & 57.96 (9.23)   & 3.97 & 3.95     \\ \hline
MovieLens 10M  & 157.53 (106.83) & 26.26 (9.71) & 176.65 (87.25)  & 5.99 & 5.71     \\ \hline
\end{tabular}
\caption{Average fitting time (and standard deviation, if available) in seconds for the \ac{MF}-based models.}
\label{tab:fitting-times}
\end{table}

It can be observed that models without reliability estimates for predictions scale much better than those that do provide them. While the former show a nearly constant fitting time regardless of dataset size, the latter experience increased fitting times as the dataset grows.

Regarding the comparison between the proposed \ac{ResBeMF} model and the \ac{BeMF} and \ac{DirMF} models, it is notable that the proposed model is faster on small datasets but slower on larger ones compared to \ac{BeMF} (keeping in mind that \ac{BeMF} produced very poor results on MovieLens 10M, as shown in \cref{fig:mf-test-error}). Additionally, the average fitting time of the proposed model is similar to that of \ac{DirMF} across all datasets except for MovieLens 1M, although the standard deviation is higher for the proposed model. These variations are mainly due to differences in the number of iterations and latent factors each model requires for optimal fitting (see \cref{fig:ml100k-hyperparameters,fig:ml1m-hyperparameters,fig:ft-hyperparameters,fig:anime-hyperparameters,fig:ml10m-hyperparameters}).

\section{Conclusions and future work} \label{sec:conclusions}

In this paper, we have introduced \ac{ResBeMF}, a new \ac{MF} model that addresses \ac{CF} as a classification problem. The output of the model is a tuple $\langle \textrm{prediction}, \textrm{reliability}\rangle$, so the unreliable predictions made by the model can be filtered out to increase the accuracy of the model. The hyper-parameter tuning has been treated as a multi-objective optimization problem, trying to discover those hyper-parameters that maximize both the quality and quantity of predictions.

The experimental results conducted demonstrate the proposed model \ac{ResBeMF}, as the model that offers a better balance between quality and predictive capability. Some state-of-the-art models, such as \ac{PMF}, \ac{BiasedMF} or \ac{MLP}, can provide a higher number of predictions at the expense of sacrificing the quality of predictions and recommendations. However, other models, such as \ac{DirMF}, \ac{GCMC} or \ac{NGCF}, provide higher quality predictions and recommendations, but, in turn, the number of predictions they can make with high reliability is lower. Furthermore, it is interesting to highlight the differences with respect to its predecessor model, \ac{BeMF}. On average, the quality and predictive capability of both models are very similar; however, the Pareto front of the \ac{ResBeMF} model is broader, allowing for a better fit of the type of \ac{RS} we want to use. Therefore, we can customize our \ac{RS} to have higher predictive quality or better predictive capability.

The techniques applied in this paper open a new horizon of possibilities regarding the use of hyper-parameters to push the Pareto front of the multi-objective optimization problem that confronts accuracy and coverage. In particular, as future work, we propose to introduce new hyper-parameters into \ac{CF} models associated with regularization coefficients of the loss function. This would allow us to have better control over the balance between quantity of predictions and quality of predictions of the model output, leading to even more flexibility of the models.

\section*{Acknowledgements}

This work was partially supported by \textit{Ministerio de Ciencia e Innovación} of Spain under the project PID2019-106493RB-I00 (DL-CEMG) and the \textit{Comunidad de Madrid} under \textit{Convenio Plurianual} with the Universidad Politécnica de Madrid in the actuation line of \textit{Programa de Excelencia para el Profesorado Universitario}.

\bibliographystyle{abbrv} 
\bibliography{refs}

@inproceedings{yang2021multi,
  title={Multi-output gaussian processes for uncertainty-aware recommender systems},
  author={Yang, Yinchong and Buettner, Florian},
  booktitle={Uncertainty in Artificial Intelligence},
  pages={1505--1514},
  year={2021},
  organization={PMLR}
}

@article{ortega2018cf4j,
  title={{CF4J}: Collaborative filtering for {J}ava},
  author={Ortega, Fernando and Zhu, Bo and Bobadilla, Jes{\'u}s and Hernando, Antonio},
  journal={Knowledge-Based Systems},
  volume={152},
  pages={94--99},
  year={2018},
  publisher={Elsevier}
}

@article{ortega2021cf4j,
  title={{CF4J} 2.0: Adapting Collaborative Filtering for {J}ava to new challenges of collaborative filtering based recommender systems},
  author={Ortega, Fernando and Mayor, Jes{\'u}s and L{\'o}pez-Fern{\'a}ndez, Daniel and Lara-Cabrera, Ra{\'u}l},
  journal={Knowledge-Based Systems},
  volume={215},
  pages={106629},
  year={2021},
  publisher={Elsevier}
}

@article{ortega2021providing,
  title={Providing reliability in recommender systems through {B}ernoulli {M}atrix {F}actorization},
  author={Ortega, Fernando and Lara-Cabrera, Ra{\'u}l and Gonz{\'a}lez-Prieto, {\'A}ngel and Bobadilla, Jes{\'u}s},
  journal={Information Sciences},
  volume={553},
  pages={110--128},
  year={2021},
  publisher={Elsevier}
}

@article{lara2022dirichlet,
  title={{D}irichlet {M}atrix {F}actorization: A Reliable Classification-Based Recommender System},
  author={Lara-Cabrera, Ra{\'u}l and Gonz{\'a}lez, {\'A}lvaro and Ortega, Fernando and Gonz{\'a}lez-Prieto, {\'A}ngel},
  journal={Applied Sciences},
  volume={12},
  number={3},
  pages={1223},
  year={2022},
  publisher={Multidisciplinary Digital Publishing Institute}
}

@article{mnih2007probabilistic,
  title={Probabilistic matrix factorization},
  author={Mnih, Andriy and Salakhutdinov, Russ R},
  journal={Advances in neural information processing systems},
  volume={20},
  year={2007}
}

@inproceedings{he2017neural,
  title={Neural collaborative filtering},
  author={He, Xiangnan and Liao, Lizi and Zhang, Hanwang and Nie, Liqiang and Hu, Xia and Chua, Tat-Seng},
  booktitle={Proceedings of the 26th international conference on world wide web},
  pages={173--182},
  year={2017}
}

@article{roetzel2019information,
  title={Information overload in the information age: a review of the literature from business administration, business psychology, and related disciplines with a bibliometric approach and framework development},
  author={Roetzel, Peter Gordon},
  journal={Business research},
  volume={12},
  number={2},
  pages={479--522},
  year={2019},
  publisher={Springer}
}

@article{resnick1997recommender,
  title={Recommender systems},
  author={Resnick, Paul and Varian, Hal R},
  journal={Communications of the ACM},
  volume={40},
  number={3},
  pages={56--58},
  year={1997},
  publisher={ACM New York, NY, USA}
}

@article{park2012literature,
  title={A literature review and classification of recommender systems research},
  author={Park, Deuk Hee and Kim, Hyea Kyeong and Choi, Il Young and Kim, Jae Kyeong},
  journal={Expert systems with applications},
  volume={39},
  number={11},
  pages={10059--10072},
  year={2012},
  publisher={Elsevier}
}

@article{bobadilla2013recommender,
  title={Recommender systems survey},
  author={Bobadilla, Jes{\'u}s and Ortega, Fernando and Hernando, Antonio and Guti{\'e}rrez, Abraham},
  journal={Knowledge-based systems},
  volume={46},
  pages={109--132},
  year={2013},
  publisher={Elsevier}
}

@article{koren2022advances,
  title={Advances in collaborative filtering},
  author={Koren, Yehuda and Rendle, Steffen and Bell, Robert},
  journal={Recommender systems handbook},
  pages={91--142},
  year={2022},
  publisher={Springer}
}

@article{kluver2018rating,
  title={Rating-based collaborative filtering: algorithms and evaluation},
  author={Kluver, Daniel and Ekstrand, Michael D and Konstan, Joseph A},
  journal={Social Information Access},
  pages={344--390},
  year={2018},
  publisher={Springer}
}

@article{ahn2008new,
  title={A new similarity measure for collaborative filtering to alleviate the new user cold-starting problem},
  author={Ahn, Hyung Jun},
  journal={Information sciences},
  volume={178},
  number={1},
  pages={37--51},
  year={2008},
  publisher={Elsevier}
}

@article{bobadilla2010new,
  title={A new collaborative filtering metric that improves the behavior of recommender systems},
  author={Bobadilla, Jes{\'u}s and Serradilla, Francisco and Bernal, Jesus},
  journal={Knowledge-Based Systems},
  volume={23},
  number={6},
  pages={520--528},
  year={2010},
  publisher={Elsevier}
}

@article{koren2009matrix,
  title={Matrix factorization techniques for recommender systems},
  author={Koren, Yehuda and Bell, Robert and Volinsky, Chris},
  journal={Computer},
  volume={42},
  number={8},
  pages={30--37},
  year={2009},
  publisher={IEEE}
}

@article{bobadilla2018reliability,
  title={Reliability quality measures for recommender systems},
  author={Bobadilla, Jesus and Guti{\'e}rrez, Abraham and Ortega, Fernando and Zhu, Bo},
  journal={Information Sciences},
  volume={442},
  pages={145--157},
  year={2018},
  publisher={Elsevier}
}

@article{katarya2018reliable,
  title={Reliable recommender system using improved collaborative filtering technique},
  author={Katarya, Rahul},
  journal={System Reliability Management: Solutions and Technologies},
  volume={113},
  pages={1556--6013},
  year={2018},
  publisher={CRC Press}
}

@article{ahmadian2019novel,
  title={A novel approach based on multi-view reliability measures to alleviate data sparsity in recommender systems},
  author={Ahmadian, Sajad and Afsharchi, Mohsen and Meghdadi, Majid},
  journal={Multimedia tools and applications},
  volume={78},
  number={13},
  pages={17763--17798},
  year={2019},
  publisher={Springer}
}

@article{zhu2018assigning,
  title={Assigning reliability values to recommendations using matrix factorization},
  author={Zhu, Bo and Ortega, Fernando and Bobadilla, Jes{\`u}s and Guti{\'e}rrez, Abraham},
  journal={Journal of computational science},
  volume={26},
  pages={165--177},
  year={2018},
  publisher={Elsevier}
}

@Article{harper2015movielens,
  author    = {Harper, F. Maxwell and Konstan, Joseph A.},
  journal   = {ACM Transactions on Interactive Intelligent Systems},
  title     = {{The movielens datasets: History and context}},
  year      = {2015},
  number    = {4},
  pages     = {1--19},
  volume    = {5},
  publisher = {ACM New York, NY, USA},
}

@InProceedings{guo2013novel,
  author    = {Guo, G. and Zhang, J. and Yorke-Smith, N.},
  booktitle = {Proceedings of the 23rd International Joint Conference on Artificial Intelligence (IJCAI)},
  title     = {{A Novel Bayesian Similarity Measure for Recommender Systems}},
  year      = {2013},
  pages     = {2619--2625},
}

@article{gorgoglione2019recommendation,
  title={Recommendation strategies in personalization applications},
  author={Gorgoglione, Michele and Panniello, Umberto and Tuzhilin, Alexander},
  journal={Information \& Management},
  volume={56},
  number={6},
  pages={103143},
  year={2019},
  publisher={Elsevier}
}

@article{wu2022graph,
  title={Graph neural networks in recommender systems: a survey},
  author={Wu, Shiwen and Sun, Fei and Zhang, Wentao and Xie, Xu and Cui, Bin},
  journal={ACM Computing Surveys},
  volume={55},
  number={5},
  pages={1--37},
  year={2022},
  publisher={ACM New York, NY}
}

@article{tang2021dynamic,
  title={Dynamic evolution of multi-graph based collaborative filtering for recommendation systems},
  author={Tang, Hao and Zhao, Guoshuai and Bu, Xuxiao and Qian, Xueming},
  journal={Knowledge-Based Systems},
  volume={228},
  pages={107251},
  year={2021},
  publisher={Elsevier}
}

@inproceedings{xia2022hypergraph,
  title={Hypergraph contrastive collaborative filtering},
  author={Xia, Lianghao and Huang, Chao and Xu, Yong and Zhao, Jiashu and Yin, Dawei and Huang, Jimmy},
  booktitle={Proceedings of the 45th International ACM SIGIR conference on research and development in information retrieval},
  pages={70--79},
  year={2022}
}

@inproceedings{rendle2020neural,
  title={Neural collaborative filtering vs. matrix factorization revisited},
  author={Rendle, Steffen and Krichene, Walid and Zhang, Li and Anderson, John},
  booktitle={Proceedings of the 14th ACM Conference on Recommender Systems},
  pages={240--248},
  year={2020}
}

@article{papadakis2022collaborative,
  title={Collaborative filtering recommender systems taxonomy},
  author={Papadakis, Harris and Papagrigoriou, Antonis and Panagiotakis, Costas and Kosmas, Eleftherios and Fragopoulou, Paraskevi},
  journal={Knowledge and Information Systems},
  volume={64},
  number={1},
  pages={35--74},
  year={2022},
  publisher={Springer}
}

@article{wu2022survey,
  title={A survey on accuracy-oriented neural recommendation: From collaborative filtering to information-rich recommendation},
  author={Wu, Le and He, Xiangnan and Wang, Xiang and Zhang, Kun and Wang, Meng},
  journal={IEEE Transactions on Knowledge and Data Engineering},
  volume={35},
  number={5},
  pages={4425--4445},
  year={2022},
  publisher={IEEE}
}

@article{berg2017graph,
  title={Graph convolutional matrix completion},
  author={Berg, Rianne van den and Kipf, Thomas N and Welling, Max},
  journal={arXiv preprint arXiv:1706.02263},
  year={2017}
}

@article{zhu2004recall,
  title={Recall, precision and average precision},
  author={Zhu, Mu},
  journal={Department of Statistics and Actuarial Science, University of Waterloo, Waterloo},
  volume={2},
  number={30},
  pages={6},
  year={2004}
}

@inproceedings{wang2019neural,
  title={Neural graph collaborative filtering},
  author={Wang, Xiang and He, Xiangnan and Wang, Meng and Feng, Fuli and Chua, Tat-Seng},
  booktitle={Proceedings of the 42nd international ACM SIGIR conference on Research and development in Information Retrieval},
  pages={165--174},
  year={2019}
}

@article{li2023trustworthy,
  title={Trustworthy recommender systems: A survey},
  author={Li, Yumeng and Zhang, Huiyan and Wang, Zhiang and Yang, Min and Zhang, Ji},
  journal={ACM Computing Surveys},
  volume={56},
  number={5},
  pages={1--38},
  year={2023},
  publisher={ACM New York, NY}
}

@inproceedings{si2023conformal,
  title={When Recommendation Systems Meet Conformal Prediction: A Survey},
  author={Si, Yaochen and Zhang, Yifan and Tsang, Ivor W},
  booktitle={Proceedings of the 32nd ACM International Conference on Information and Knowledge Management},
  pages={1--10},
  year={2023}
}

@inproceedings{wang2023diffusion,
  title={Diffusion recommender model},
  author={Wang, Wenjie and Feng, Fuli and He, Xiangnan and Zhang, Hanwang and Chua, Tat-Seng},
  booktitle={Proceedings of the 46th International ACM SIGIR Conference on Research and Development in Information Retrieval},
  pages={832--841},
  year={2023}
}

\appendix

\section{Random Search Results}
\label{sec:appendix}

\begin{figure}[H]
    \centering
    \includegraphics[width=0.95\textwidth]{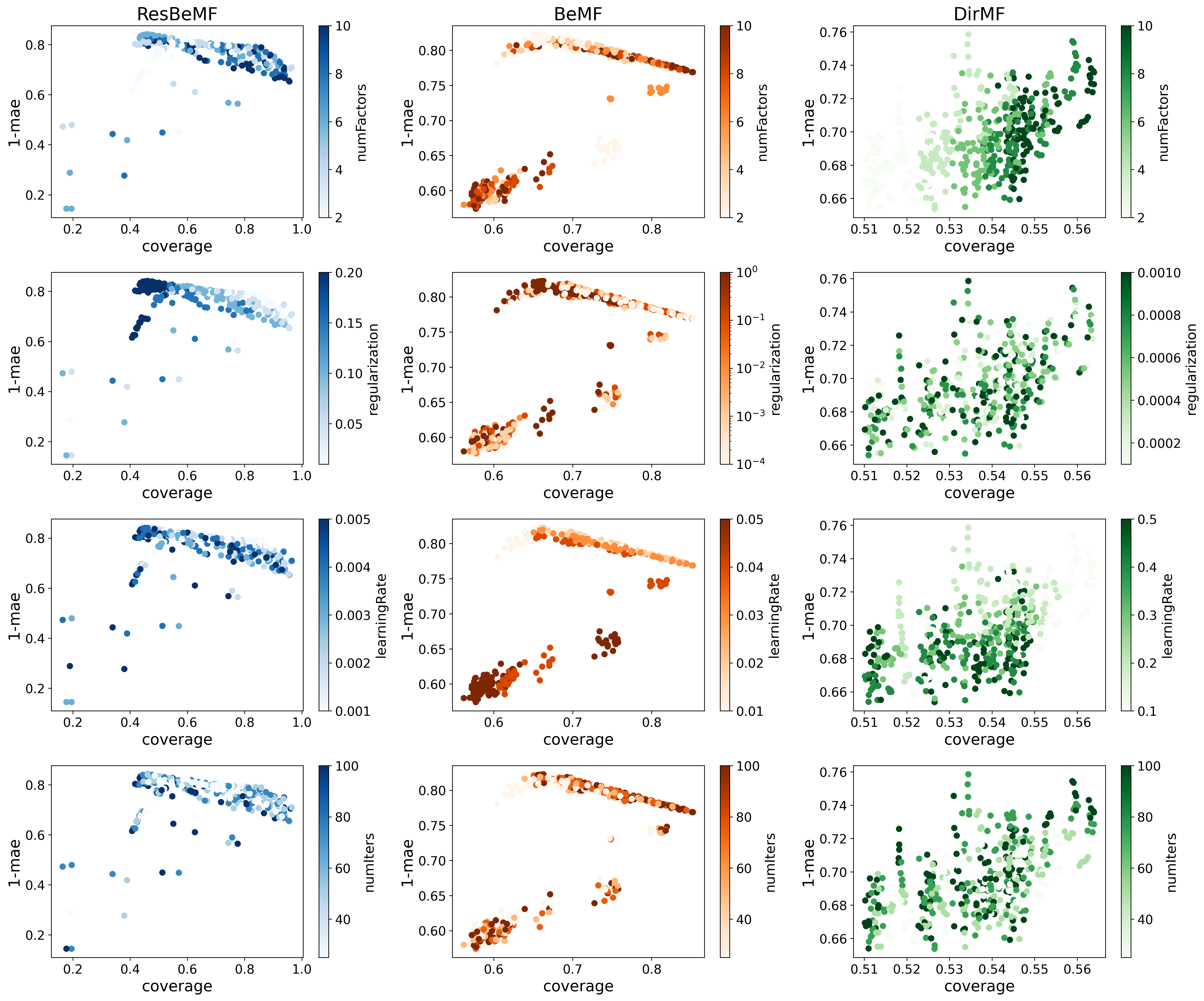}
    \caption{Random search results for parameter tuning in MovieLens 100K dataset.}
    \label{fig:ml100k-hyperparameters}
\end{figure}

\begin{figure}[H]
    \centering
    \includegraphics[width=0.95\textwidth]{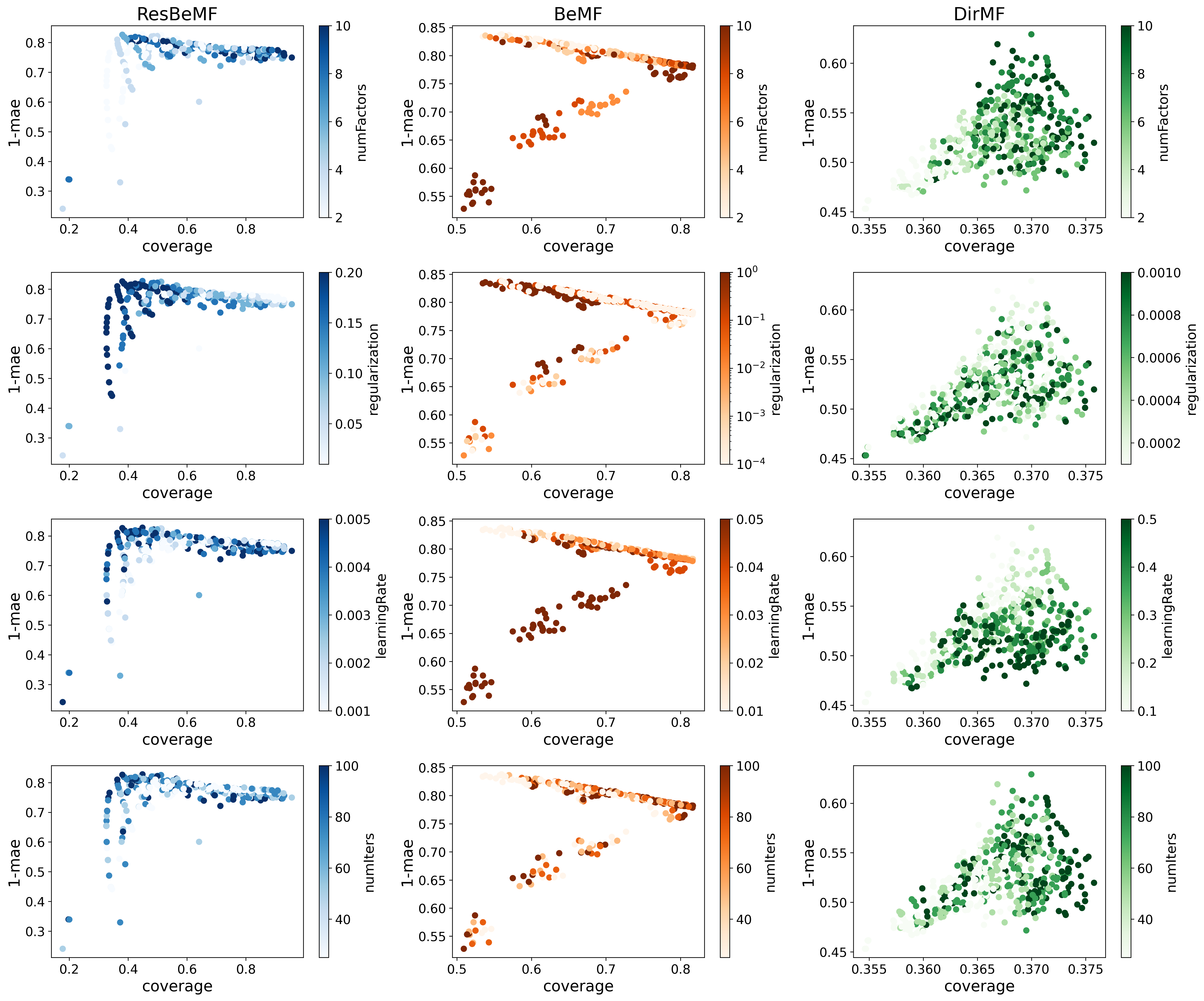}
    \caption{Random search results for parameter tuning in FilmTrust dataset.}
    \label{fig:ft-hyperparameters}
\end{figure}

\begin{figure}[H]
    \centering
    \includegraphics[width=0.95\textwidth]{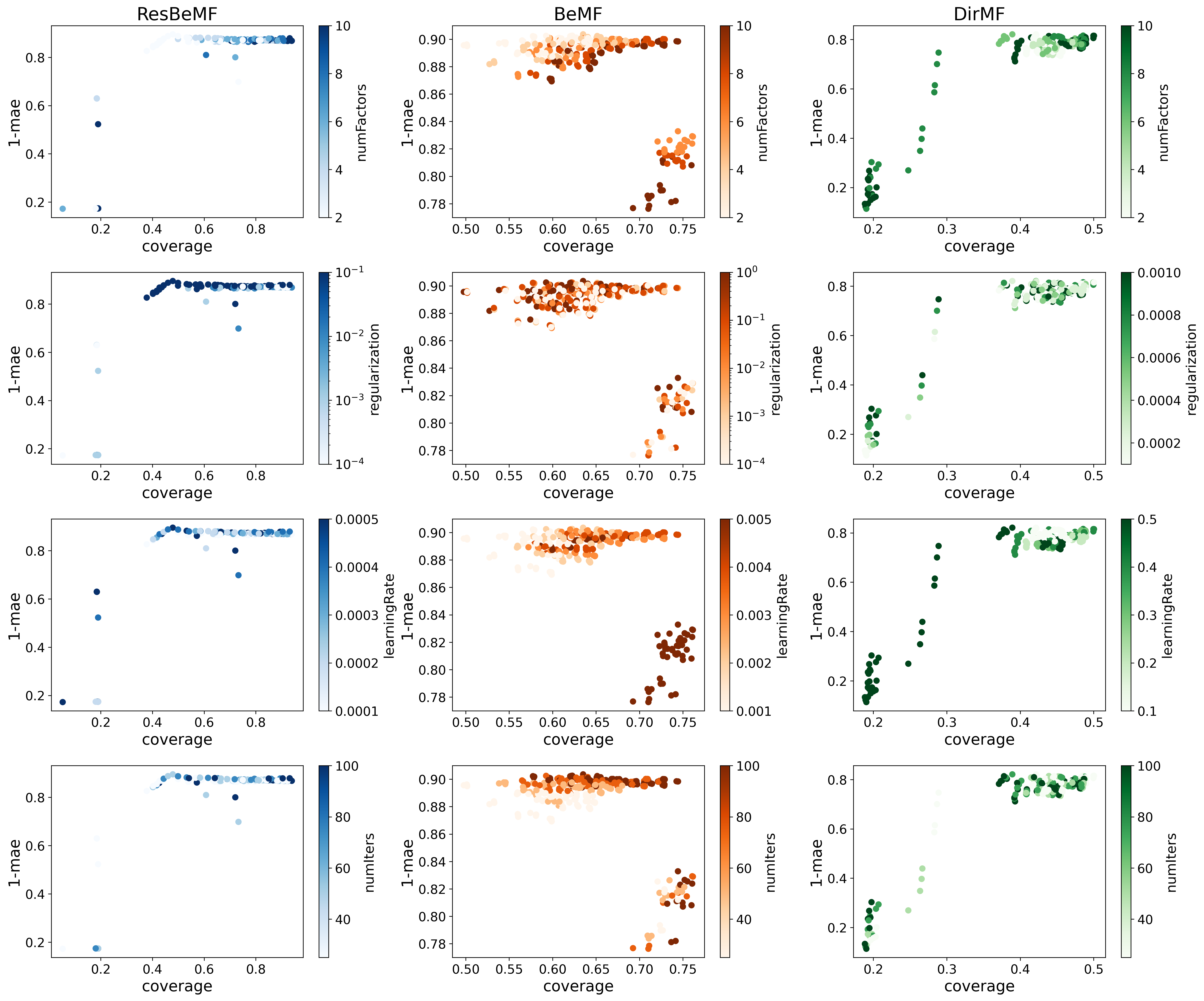}
    \caption{Random search results for parameter tuning in MyAnimeList dataset.}
    \label{fig:anime-hyperparameters}
\end{figure}

\begin{figure}[H]
    \centering
    \includegraphics[width=0.95\textwidth]{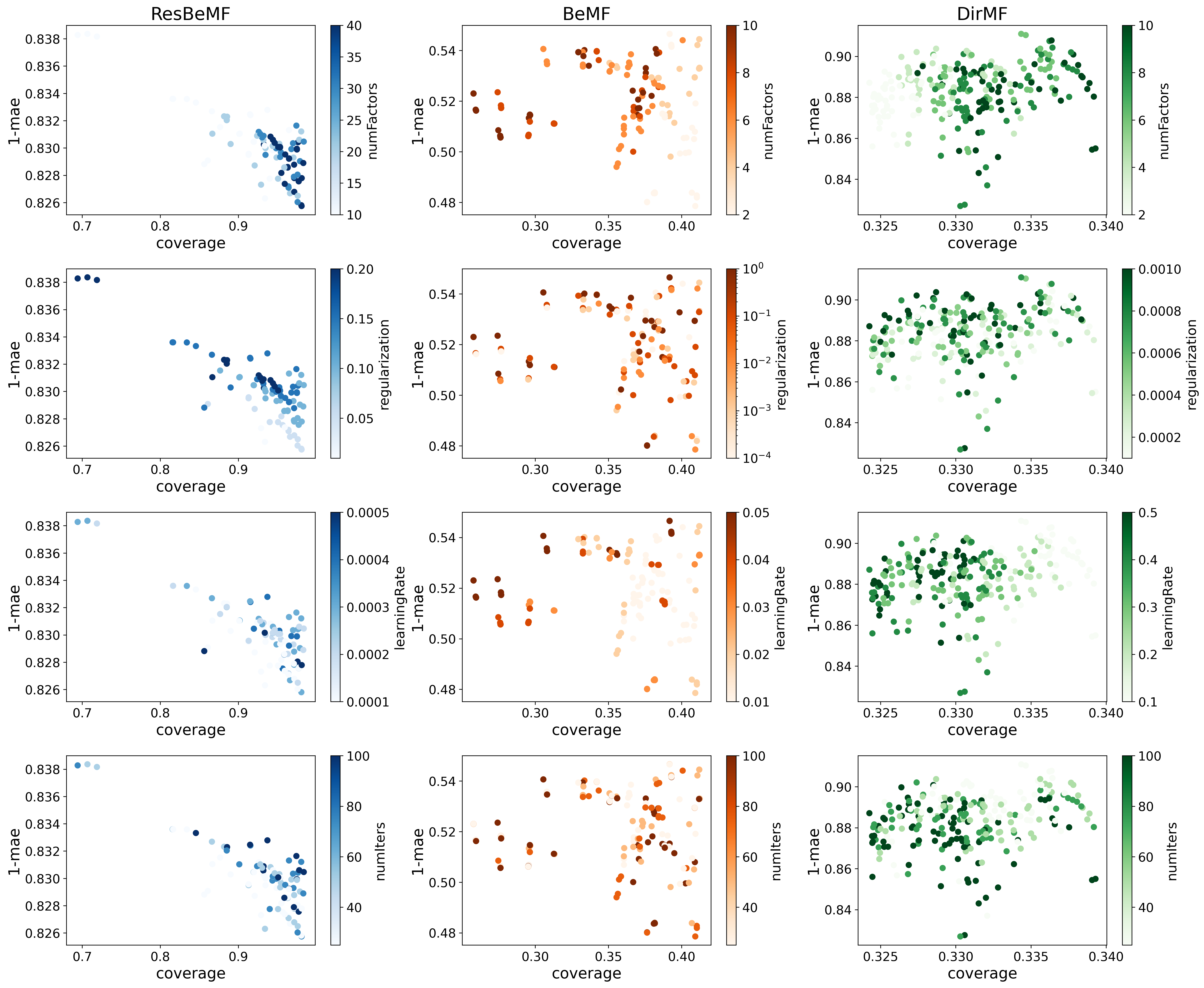}
    \caption{Random search results for parameter tuning in MovieLens10M dataset.}
    \label{fig:ml10m-hyperparameters}
\end{figure}

This appendix contains the random search results for the MovieLens100K (\cref{fig:ml100k-hyperparameters}), FilmTrust (\cref{fig:ft-hyperparameters}), MyAnimeList (\cref{fig:anime-hyperparameters}), and MovieLens 10M (\cref{fig:ml10m-hyperparameters}) datasets.

\end{document}